\documentclass[ pra,twocolumn,superscriptaddress]{revtex4}
\usepackage[colorlinks=true, urlcolor=blue, pdfborder={0 0 0}]{hyperref}

\usepackage{graphicx}
\usepackage{amsmath}
\usepackage{mathrsfs}
\usepackage{amsfonts}

\usepackage{subfigure}
\usepackage{epsfig} 
\usepackage{epstopdf}
\usepackage{float} 
\usepackage{tikz}
\usetikzlibrary{shapes.arrows}
\usepackage[percent]{overpic}
\usepackage{comment}
\usepackage{pstricks}

\usepackage[utf8]{inputenc}

\begin{document}

\title{\bf A Generalization of Fermat's Principle for Classical and Quantum Systems}
\author{Tarek A. Elsayed}
\email{T.Elsayed@thphys.uni-heidelberg.de}
\address{Institute of Theoretical Physics, University of Heidelberg, Philosophenweg 19, 69120 Heidelberg, Germany}

\begin{abstract}

The analogy between dynamics and optics had a great influence on the development of the foundations of classical and quantum mechanics. We take this analogy one step further and investigate the validity of Fermat's principle in many-dimensional spaces describing dynamical systems (i.e., the quantum Hilbert space and the classical phase and configuration space). We propose that if the notion of a metric distance is well defined in that space and the velocity of the representative point of the system is an invariant of motion, then a generalized version of Fermat's principle will hold. We substantiate this conjecture for time-independent quantum systems and for a classical system consisting of coupled harmonic oscillators. An exception to this principle is the configuration space of a charged particle in a constant magnetic field; in this case the principle is valid in a frame rotating by half the Larmor frequency, not the stationary lab frame. 

\end{abstract}
\maketitle

\section{Introduction}
  An important lesson that has been emphasized throughout the history of physics is that illuminating new aspects of the interwoven connections between  geometry  and physics leads to paradigm shifts in physics. Typically, novel geometric considerations of physical quantities lead to new variational principles which assign the natural evolution of physical systems with an extremum of some functional or a geodesic curve in some hyperspace. The oldest of these variational principles is the Fermat principle of least time, which became a fundamental principle in geometric optics. The principle was introduced by Fermat, who also called it \emph{ the principle of natural economy} \cite{basdevant}, and it states that light rays travel in a general  medium along the path that minimizes the time taken to travel between the initial and final destinations. The concept of natural economy inspired Maupertuis to introduce the principle of least action in analytical mechanics, which later evolved through the work of Euler, Lagrange, Hamilton, and Jacobi to become a fundamental concept in classical mechanics. By 1887, it had become clear that the least action is a universal concept in physics when Helmholtz expanded its domain of validity by applying it to two regimes beyond the standard problems of classical mechanics, namely, thermodynamics and electrodynamics \cite{helmholtz}. Since then, the pursuit of new variational principles in physics has not relented \cite{novikov2004}.
  
  The mathematical formulation of Fermat's principle states that the time functional $\mathcal{T}$, defined as
  \begin{equation}
\mathcal{T}=\int\frac{ds}{\nu(s)},
\label{T}
\end{equation}
where $ \nu(s) $ is the speed of light and $ ds $ is the distance element along the light trajectory, is minimized \cite{mandelstam}. In other words, if  $ \mathcal{T} $ is computed along all possible trajectories between fixed initial and final positions, $ \mathcal{T} $ will always be minimum along the actual path traveled by the light rays (the physical path). The modern version of Fermat's principle is written in terms of the index of refraction $ n(s) =\frac{c}{\nu(s)} $, where c is the speed of light in  free space, and states that the optical path length  $\int n(s)ds$ is a minimum. In that sense, Fermat's principle is the optical analog of Jacobi's principle of least action  \cite{lanczos}, which states that for a conservative classical system at energy $ E $, with potential function $V$ between its constituent particles, the action functional \begin{equation}
I=\int\sqrt{E-V(s)}ds
\label{jac}
\end{equation}
is an extremum. 

The remarkable property of this action that distinguishes it from other variational principles in analytical mechanics, i.e., Hamilton and Lagrange's variational principles,  is that it represents a purely geometric quantity. This quantity is computed along different trajectories in the configuration space between fixed points without referring to any time evolution. Therefore, Eq. (\ref{jac}) can be used to define a new Riemannian space, whose metric $ ds'=\sqrt{E-V(s)}ds $, where the natural evolution of the representative point of the system is along geodesic curves.
    
 In this work, we set out to seek how far the analogy between dynamics and optics applies as far as the Fermat principle is concerned. In particular, we investigate the validity of Fermat's principle for a generic many-body classical and quantum system, and pose the following question: If the state of  a conservative dynamical system is represented in some metric space $ \mathbb{S} $ by a point, and the velocity field $ \nu(s) $ is computed everywhere in $ \mathbb{S} $ from the equations of motion using the proper metric of that space, will the motion of this point be along a path that extremizes the time functional $\mathcal{T}$? 
 
 \section{The Generalized Fermat Principle}
  We answer the question posed above by proposing the \emph{generalized fermat principle} (GFP): {Whenever the speed of the representative point of a conservative dynamical system, $\nu(s)$, is an integral of motion in a metric space  $ \mathbb{S} $,  the path followed during the dynamical evolution of that system in $ \mathbb{S} $ between fixed initial and final states makes the time functional $\mathcal{T}$ stationary against small variations of the path. } In contrast to light rays, where $\mathcal{T}$ is an extremum even when the speed of light is not constant (i.e., in an inhomogeneous medium), this conjecture considers only the case when  $\nu(s)$ is invariant during the time evolution.  A corollary that follows from this conjecture is that the length of the physical path $\int ds$ is stationary  (e.g., the path can be a geodesic) on the sub-manifold of a given value of $\nu(s)$ embedded in $\mathbb{S} $ when the above condition is fulfilled.

Mathematically speaking,  the GFP states that if $\nu(s)$ is a constant of motion along the physical path (not necessarily in the whole space), then  among all possible trajectories between initial and final states, only those which make $ \mathcal{T} $ invariant under an infinitesimal variation of the path, i.e.,
 \begin{equation}
\delta  \mathcal{T} =0,
\label{var}
\end{equation}
are possible candidates for the dynamical evolution.  The value of $ \mathcal{T} $ corresponding to the physical path is not necessarily the global minimum between all paths connecting the initial and final states. We emphasize here that we are not aiming to derive the equations of motion from the time action, because we have to use them to find $ \nu(s) $ in the first place.  We  rather propose that they necessarily lead to a stationary time action when $\nu(s)$ is an integral of motion. Unlike the original Fermat principle,  not every pair of states are connected by a physical path.  Rather, the principle proposed here gives a geometrically appealing argument to explain why the evolution of the system followed a certain trajectory between a given pair of initial and final states which we know a priori are connected by some physical path.

  We investigate the validity of this conjecture by considering three cases: (i) The evolution of quantum systems in the projective Hilbert space $\mathbb{P}$, where wavefunctions are defined up to an overall phase factor. (ii) The evolution of a system of coupled harmonic oscillators in the phase space consisting of coordinates and momenta and equipped with an Euclidean metric. (iii) The motion of a charged particle in a constant magnetic field, which turns out to be an exception. Similar to the Jacobi's principle, the principle proposed here represents a geometric variational principle in the phase and Hilbert space. In both cases, the velocity field $\nu(s)$ is defined completely by the Hamiltonian of the problem, and  is obtained from the equations of motion of the system that will drive its evolution along the physical path (i.e., Schr\"odinger equation in quantum systems and Hamilton's equations of motion in classical systems).

    \subsection{Generalized Fermat Principle in Hilbert Space}
The development of the concept of geometric phase in quantum mechanics triggered the interest of many physicists to look for more connections between quantum mechanics and geometry \cite{ashtekar}.   Anandan and Aharonov  investigated the nature of the geometry of quantum evolution in the projective Hilbert space $\mathbb{P}$ through a series of papers in the late 80s \cite{aharonov87,aharonov90,aharonov88}.  They have shown \cite{aharonov90} that the speed of quantum evolution in $\mathbb{P}$ is related to the energy uncertainty $ \Delta E=\left ( \left \langle H^2 \right \rangle - \left \langle H \right \rangle^2\right )^{1/2} $ via 
  \begin{equation}
ds= \Delta E\ dt/\hbar,
\end{equation}
where $ ds $ is the infinitesimal distance in $\mathbb{P}$ given by the Fubini-Study (FS) metric  $ ds^2= \frac{\left \langle \delta \psi  |\delta \psi  \right \rangle}{\left \langle \psi |\psi  \right \rangle} - \frac{|\left \langle \delta \psi | \psi  \right \rangle|^2}{\left \langle \psi |\psi  \right \rangle^2}$.  On the unit sphere, $ds= \langle \delta \psi |1-\hat{P}|\delta \psi\rangle^\frac{1}{2}$,  where $ \hat{P} $ is the projection operator $|\psi\rangle \langle \psi|  $. The trajectory traversed by a ray in $\mathbb{P}$ under unitary evolution is generally not a geodesic, i.e., $ \delta\int ds\neq 0  $. This can be easily conceived by considering a system composed of a single quantum spin-1/2. In this case, $\mathbb{P}$ is simply the Bloch sphere and the precession motion of the spin on Bloch sphere off the equator is not a geodesic.  

Several attempts \cite{kryukov,grigorenko} have been made to find new formulations where the quantum evolution is a geodesic flow. Noticing that the speed of quantum evolution $ \Delta E/\hbar$ is invariant for time-independent Hamiltonians,  the simple answer to this problem suggested by the present paper is to consider $ \mathcal{T}=\int \frac{ds}{\Delta E} $ as a geodesic quantity, i.e., \emph{Fermat's principle in Hilbert space}. This issue should be distinguished from the quantum brachistochrone problem \cite{optimal}, where the Hamiltonian that leads to optimal time evolution between an initial and final state is sought. The above proposition, however, states that the unitary evolution generated by any time-independent Hamiltonian is optimal, with respect to all other possible trajectories connecting the initial and final states (Fig. 1-a). 

 To show that $\mathcal{T}$ is stationary along the physical path through $\mathbb{P}$, let us parameterize the evolution along any path connecting the initial and final states $ | \psi_i  \rangle $ and $ | \psi_f  \rangle $ by some arbitrary parameter $ \tau $. We can write Eq. (\ref{T}) as 
\begin{equation}
\mathcal{T}=\int_{|\psi_i \rangle }^{|\psi_f \rangle} d\tau\frac{\langle\dot{\psi}|1-\hat{P}|\dot{\psi}\rangle^{\frac{1}{2}}}{\left ( \left \langle \psi| H^2 |\psi\right \rangle - \left \langle\psi| H|\psi \right \rangle^2\right )^{\frac{1}{2}}},
\label{TQ}
\end{equation}
where $  | \dot{\psi}  \rangle =\frac{|\delta \psi \rangle}{\delta \tau}  $.
Taking the variational derivative of  $ \mathcal{T}$ with respect to $ \langle \delta \psi| $ subject to the constraints of normalization and fixed initial and final states, we arrive at the Euler-Lagrange (EL) equation,
\begin{equation}
\frac{\delta L}{\langle \delta \psi |}-\frac{d}{d\tau}\frac{\delta L}{\langle \delta \dot{\psi}|}=0,
\label{EL}
\end{equation}
where $L$ is the integrand in Eq. (\ref{TQ}) added to the Lagrange multiplier term $\lambda(\tau) (\langle \psi |\psi \rangle-1)$. Although we have already imposed  the normalization condition in the Fubini-Study metric, we use the Lagrange-multiplier method here to ensure that the variations of the path will respect the conservation of the norm of $|\psi \rangle$.
Calling the numerator and denominator  in Eq. (\ref{TQ}), $A$ and $B$ respectively,  Eq. (\ref{EL}) reads
\begin{widetext}
\footnotesize
\begin{equation}
\begin{split}
&\left [ \frac{-1}{2AB}\langle\dot{\psi}|\psi\rangle |\dot \psi\rangle -  \frac{A}{2B^3}\left ( H^2 |\psi\rangle -2\langle H\rangle H | \psi \rangle \right ) \right ]  -\frac{1}{2AB}\left [ |\ddot\psi \rangle - \langle\psi |\dot\psi \rangle |\dot\psi \rangle -\left ( \langle \dot \psi |\dot \psi \rangle + \langle \psi|\ddot \psi \rangle \right )| \psi \rangle\right ] -\frac{1}{2}\left ( |\dot\psi \rangle - \langle\psi |\dot\psi \rangle |\psi \rangle  \right)*\\ & \left [ \frac{-1}{2AB^3}\left ( \langle \psi H^2 | \dot \psi \rangle + \langle \dot \psi | H^2|  \psi \rangle -2 \langle H \rangle\left [ \langle \dot \psi | H |\psi \rangle + \langle \psi |H|\dot\psi \rangle \right ] \right ) -  \frac{1}{2A^3B}\left ( \langle  \ddot \psi |\dot \psi \rangle - \langle \psi | \dot \psi \rangle  \left ( \langle \ddot \psi | \psi \rangle +\langle \dot \psi | \dot \psi \rangle\right ) +c.c  \right )\right ] +\\ &\lambda(\tau) |\psi \rangle =0.
\end{split}
\label{long}
\end{equation}
\normalsize
\end{widetext}
Although Eq. (\ref{long}) is a highly nonlinear equation, it is easy to verify that the Schr\"odinger equation $ |\dot{\psi}\rangle=\pm iH|\psi\rangle $ satisfies this equation with a vanishing Lagrange multiplier, and therefore makes the time functional stationary when $\tau$ equals the real time $t$. The sign ambiguity can be considered a reminiscence of the non-unique mapping between $\tau$ and $t$. In cases where quantum ergodicity applies, i.e., when ``all states within a given energy range can be reached from all other states within the range" \cite{haar}, the opposite sign can be related to the other route to reach $|\psi_f \rangle $ starting from $|\psi_i \rangle $, i.e., backward in time. 
\twocolumngrid
The above discussion provokes several interesting issues.  First, it is intriguing to explore whether there is a nonlinear Schr\"odinger equation that would satisfy Eq. (\ref{long}) with a vanishing Lagrange multiplier and fulfill the Fermat principle as a possible extension to quantum mechanics. It is unlikely that such an equation exists since  any equation that satisfies Eq. (\ref{long})  and keeps the norm of $|\psi\rangle$ conserved should also satisfy 
\begin{equation}
\frac{\langle\dot{\psi}|\dot{\psi}\rangle}{|\langle\psi|\dot{\psi}\rangle|^2}= \frac{\left \langle \psi| H^2 |\psi\right \rangle}{\left \langle\psi| H|\psi \right \rangle^2}.
\end{equation}
This equation results after taking the inner product of Eq. (\ref{long}) with $\langle\psi|$ and making use of the normalization condition of the wavefunction. Second, it has to be emphasized that Eq. (\ref{EL}) is satisfied only for time-independent Hamiltonians. A very interesting problem is how to generalize this concept to time-dependent Hamiltonians as we shall do  in the classical domain below.  Finally, we expect the Fermat principle to be equally valid for the unitary evolution of a density matrix with the FS metric replaced by the Hilbert-Schmidt metric. It would be interesting, though, to investigate whether the non-unitary evolution of the density matrix of an open quantum system described by a master equation follows a Fermat principle. We present no further details on these issues in the present paper.  


 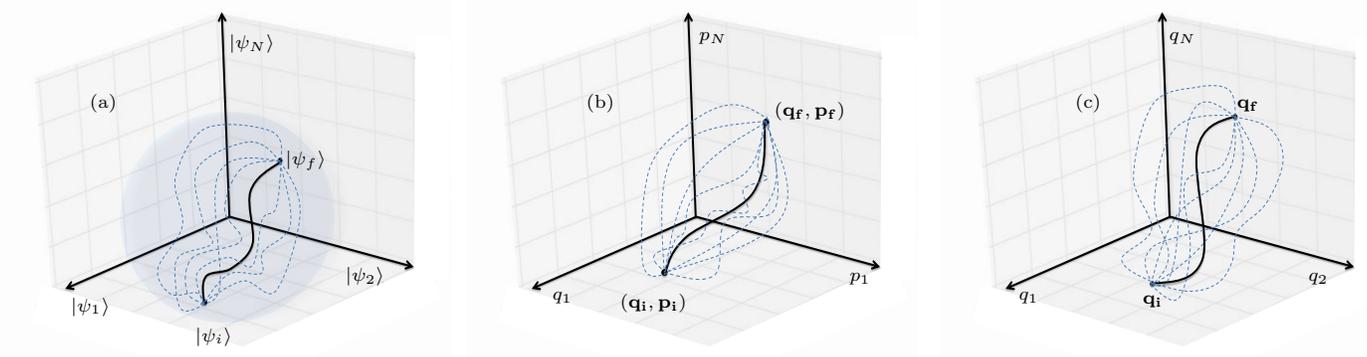
\begin{figure*}[] \setlength{\unitlength}{0.1cm}

\scriptsize
\begin{picture}(160 , 45 ) 
{
\put(-11, 0)  {\includegraphics[ width=6cm,height=5cm]{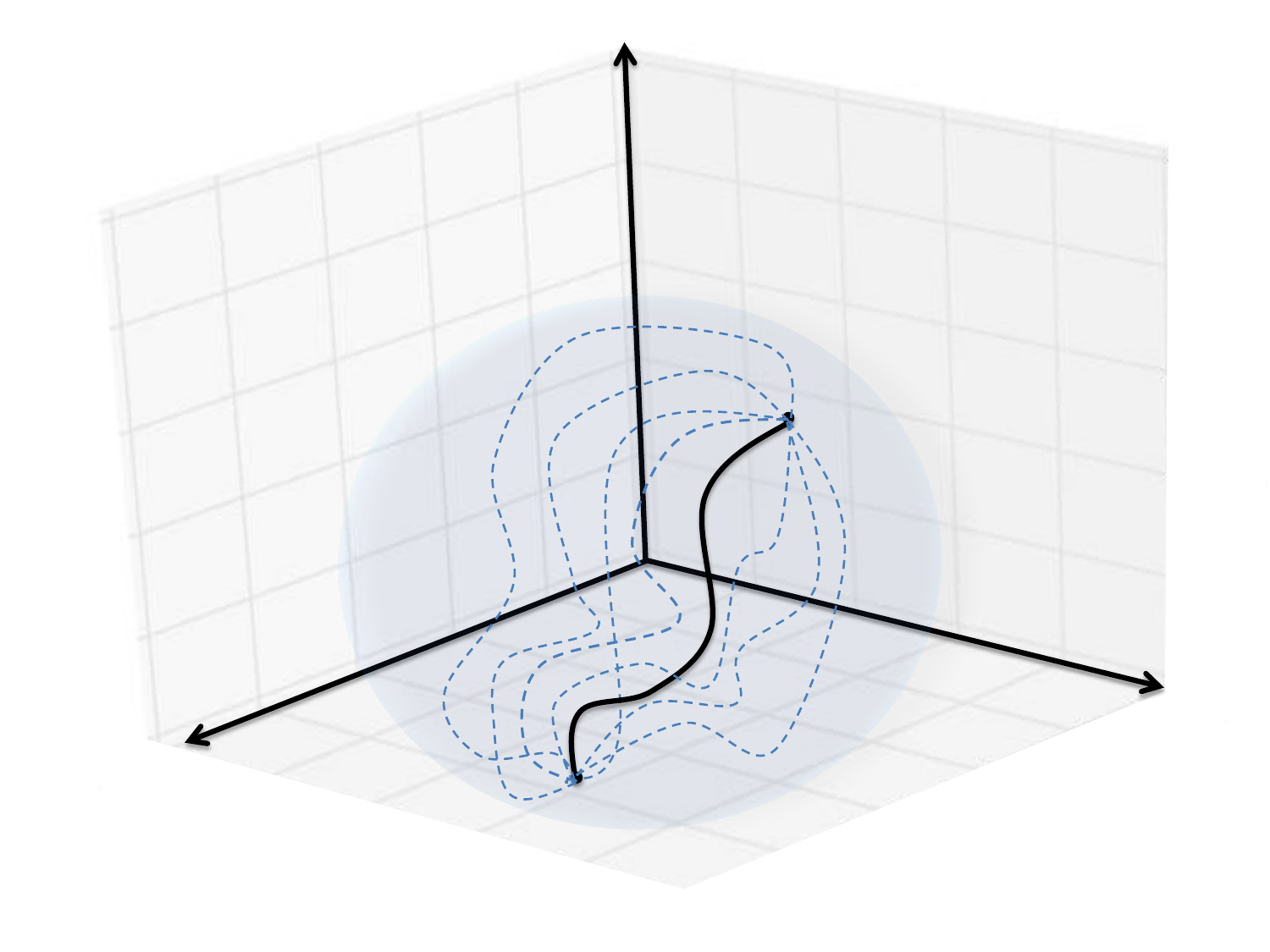}}
\put(51, 0)  {\includegraphics[ width=6cm,height=5cm]{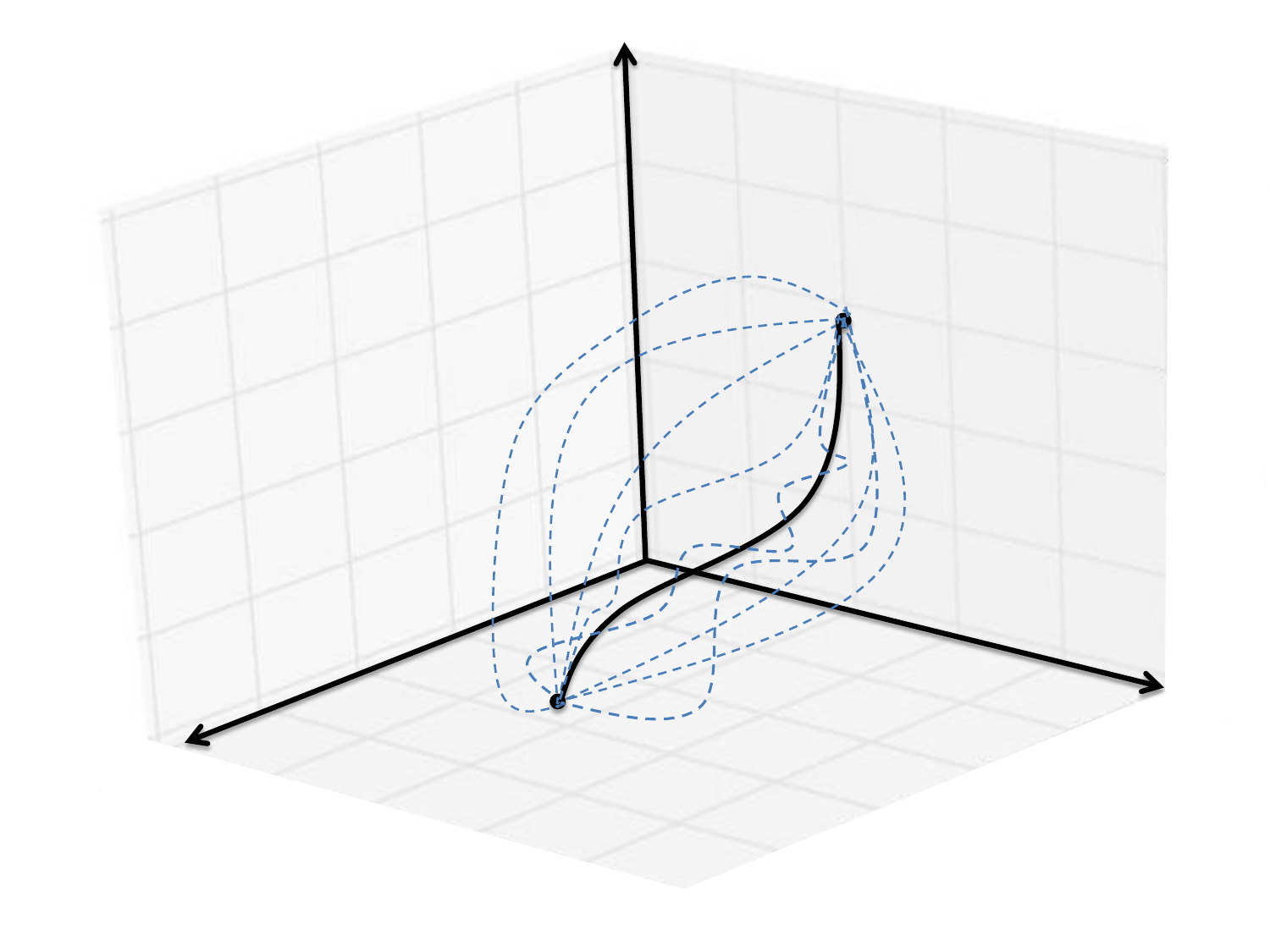}}
\put(114, 0)  {\includegraphics[ width=6cm,height=5cm]{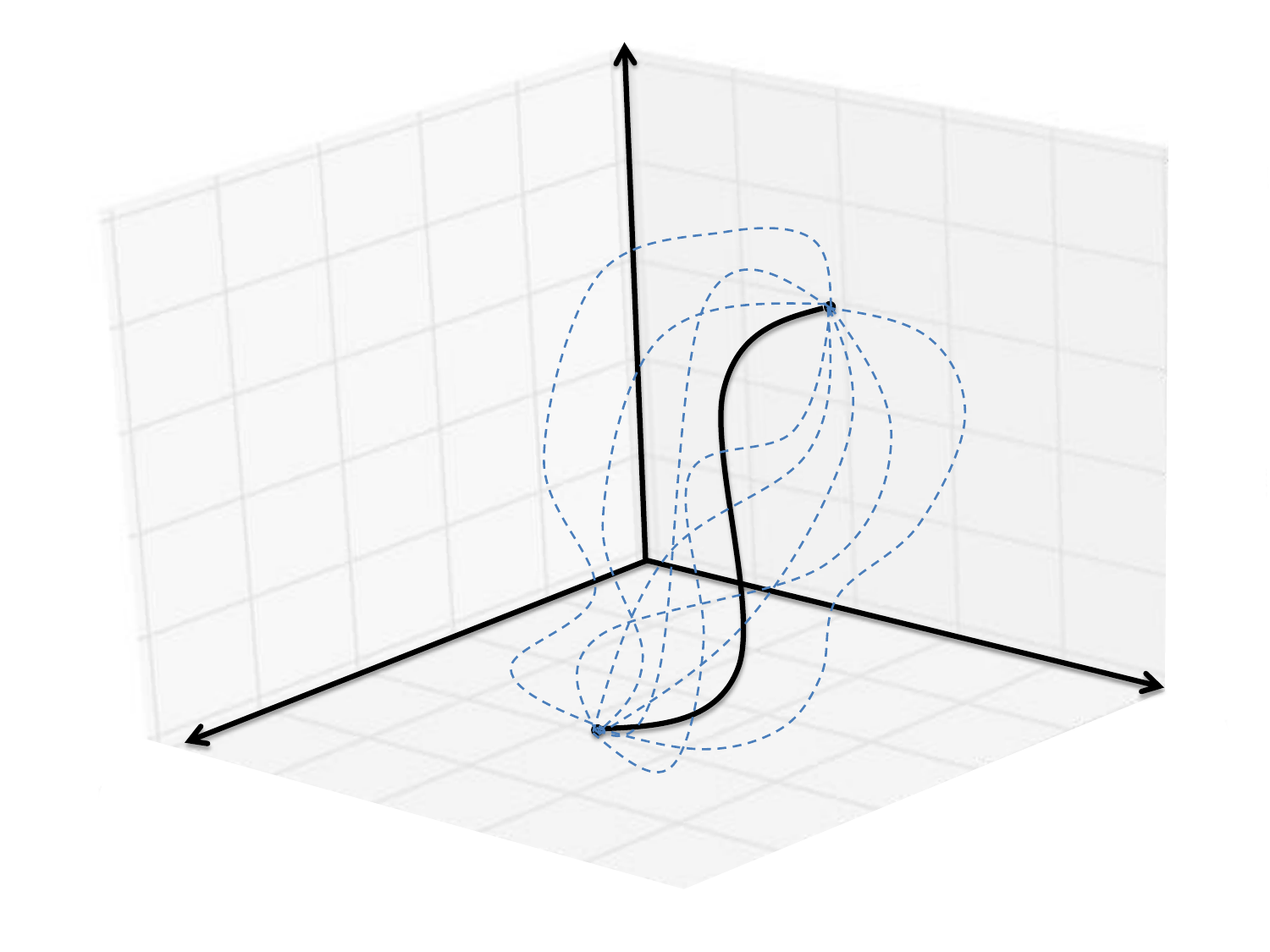}}

\put(-2.5,8) { $|\psi_1\rangle$ }
\put(34,12) { $|\psi_2\rangle$ }
\put(18.5,43) { $|\psi_N\rangle$ }

\put(14,4) { $|\psi_i\rangle$ }
\put(26,27.5) { $|\psi_f\rangle$ }

\put(61.5,9.5) { $q_1$ }
\put(101,12) { $p_1$ }
\put(81,44) { $p_N$ }

\put(70.5,8.5) { $(\mathbf{q_i},\mathbf{p_i})$ }
\put(91,34) { $(\mathbf{q_f},\mathbf{p_f})$ }

\put(123.5,9.5) { $q_1$ }
\put(162,12) { $q_2$ }
\put(143.5,44) { $q_N$ }

\put(140,9) { $\mathbf{q_i}$ }
\put(152.5,35) { $\mathbf{q_f}$ }

\put(0,35) { (a) }
\put(66,35) { (b) }
\put(131,35) { (c) }
}
\end{picture} 

\caption{ \label{fig} Natural evolution of a physical system from an initial to a final state (thick) on the unit sphere in the projective Hilbert space  (a), in the classical phase space (b) and in the classical configuration space (c) together with fictitious paths (dashed) connecting the same states. The time functional $\mathcal{T}$ is stationary when the speed of evolution defined by the Hamiltonian is constant along the physical path. This condition is satisfied naturally for time independent Hamiltonians in case (a) and only in special cases of (b) and (c) (see text).}
\end{figure*} 
 \normalsize

\subsection{Generalized Fermat Principle in Phase Space}
In optics, the Euler-Lagrange equation for the functional $\frac{1}{c}\int n(s)ds$  reduces to the ray equation (also called the eikonal equation) \cite{johns2011} 
\begin{equation}
\frac{d}{ds} (n(s) \mathbf{\hat{t}}) -\mathbf{\nabla} n(s)=0,
\label{EL-n}
\end{equation}
where $\mathbf{\hat{t}}$ is a unit tangent vector defined in terms of the position vector $\mathbf{r}$  as $\mathbf{\hat{t}}=d\vec {\mathbf{r}}/ds$ (the length of the path $s$ plays the role of time in this derivation). When $n(s)$ is constant along the path (i.e., independent of $s$), Eq. (\ref{EL-n}) can be rewritten as
\begin{equation}
\frac{d}{ds} ( \mathbf{\hat{t}})=-\frac{\mathbf{\nabla}\nu(s)}{\nu(s)}.
\label{curv}
\end{equation}
The left hand side of this equation represents a curvature vector $ \vec{\mathbf{\kappa}} $ whose magnitude equals the curvature of the path and direction is orthogonal to the direction of motion. Therefore, for the case of an Euclidean space $\mathbb{S}$, the GFP is equivalent to stating that $ \vec{\mathbf{\kappa}} $ equals $-\mathbf{\nabla}\nu(s)/ \nu(s)$ for a dynamical system that has invariable speed $\nu(s)$  along its evolution in $ \mathbb{S} $. On the other hand, since $ \vec{\mathbf{\kappa}} $ is the acceleration vector of the representative point of the system, we can regard the RHS of Eq. (\ref{curv}) as a force that drives its evolution. The potential function responsible for this force is $\log(\nu(s))$.

 We now consider a generic conservative classical system composed of $ N $ particles described by a set of generalized coordinates $ \{q_i,p_i\} $ that have the same units and Hamiltonian $H$ (Fig. 1-b).  Let the distance element in the phase space be described by the Euclidean metric $ ds^2=\sum_{i=1}^{N}dq_i^2+dp_i^2 $. The speed $ \nu(s) $ along the physical path generated by the Hamiltonian flow equals $ \sqrt{\sum_{i=1}^N \left( \frac{\partial H}{\partial q_i}\right)^2+\left(\frac{\partial H}{\partial p_i}\right)^2} $ \cite{haar}. We therefore express the time functional $\mathcal{T}$ as
 \begin{equation}
\mathcal{T}=\int_{\vec{\mathbf{x}}_i}^{\vec{\mathbf{x}}_f}\frac{ds}{\sqrt{\sum_i \left( \frac{\partial H}{\partial x_i}\right)^2}},
\end{equation} 
where $x_i$ denotes any of the generalized coordinates $ {q_i,p_i} $ treated on an equal footing and $ \vec{\mathbf{x}}$ denotes $\{ x_1,x_2,...x_{2N} \}$.

   If we parameterize any arbitrary trajectory connecting $ \vec{\mathbf{x}}_i $ and $ \vec{\mathbf{x}}_f $ by $ \tau $, then  the EL equation which satisfies the variational principle $\delta \mathcal{T} =0$ is
\begin{equation}
\frac{\delta L}{\delta x }-\frac{d}{d\tau}\frac{\delta L}{\delta \dot{x} }=0,
\label{ELPS}
\end{equation}
where $ L= FG$,  $ F=\frac{1}{\sqrt{\sum_i\left(\frac{\partial H}{\partial x_i}\right)^2}} $, $G=\sqrt{\sum_i \dot{x_i}^2}$ and $x$ denotes any of the $2N$ coordinates. The explicit form of Eq. (\ref{ELPS}) for coordinate $x_i$ is

\begin{equation}
G^2\frac{\partial F}{\partial x_i}-\frac{G^2\left(\ddot{x_i}F+\dot{x_i}\sum_{j} \frac{\partial F}{\partial x_j}\dot{x_j}\right )-\dot{x_i}F\sum_j\dot{x_j}\ddot{x_j}}{G^2}=0.
\label{fullELPS}
\end{equation}

This equation is not generally satisfied for an arbitrary choice of the generalized coordinates ${q_i,p_i}$. The only system known to the author for which a metric space is defined and $\nu(s)$ is a constant of motion is the harmonic oscillator. It is easy to verify numerically that Eq. (\ref{fullELPS}) is satisfied for a system of $N$ coupled harmonic oscillators with equal masses and coupling constants with Hamiltonian $H=\sum_{i} \frac{p_i^2}{2m}+\sum_{i<j}\frac{1}{2}m\omega^2(q_i-q_j)^2$ if we work in the phase space  $\{P_i,Q_i\}$, where $P_i=p_i$ and $Q_i=m\omega\sqrt{N} q_i$. The speed of evolution $\nu(s)$ in this case is proportional to the energy. In this example, we had to perform a simple scale transformation in order to obtain a new set of  coordinates that have the same units, leading to a metric space in which Fermat's principle is fulfilled. In other examples, one might need to look for more complicated  transformations that would render $\nu(s)$  an integral of motion in the new phase space. 

 For externally driven systems whose time-dependent  Hamiltonian is $H(p,q,t)$, we define the new Hamiltonian, $H'=H(q,p,\tau) +\eta$, where $(\tau,\eta)$ is an extra degree of freedom that satisfies $\dot{\tau}=\partial H'/\partial \eta=1$ and $\dot{\eta}=-\partial H'/\partial \tau=-\partial H/\partial \tau$ \cite{hossein}. The above discussion can be generalized to the extended phase space which consists of the original phase space supplemented by the new coordinates  $(\tau,\eta)$.

\subsection{Generalized Fermat Principle in Configuration Space}
An obvious example where GFP is valid in configuration space is the free motion of a single particle constrained to a curved surface following a geodesic trajectory. Similarly, for a system of free particles, GFP holds in the configuration space $\{q_i\}$ (Fig. 1-c), while for a system of interacting particles it holds in the Riemannian manifold defined by the metric $ds'$ introduced in the introduction. Another example is the bound states of the two-body Kepler problem. The 3D motion of the reduced one-body problem can be mapped onto the motion of a free particle on the inner surface of an $S^3$ sphere embedded in a 4D space \cite{moser,guillemin}. In the rest of this section, we discuss the applicability of Fermat's principle in the configuration space of another system, namely a charge $q$ with mass $m$ moving in a constant uniform magnetic field $\mathbf{B}$.

 The dynamics of the charge controlled by the Lorentz Force $m \frac{d\mathbf{v}}{dt}=q\mathbf{v}\times \mathbf{B}$ guarantees that the magnitude of the charge velocity is constant and GFP indicates that the time functional should be stationary in the configuration space of the charge. However, in this case, a stationary time functional $\mathcal{T}$ indicates that the length of the path of the charge in space is an extremum, which we know is not the case; the circular path of the charge in a constant magnetic field is not an  extremal path. Therefore, we consider this case a clear counter-example to the GFP if we do not restrict it to velocity-independent potentials. However, we note that even in this case, there is another frame of reference where GFP is valid. What we will show next is that in a  frame rotating with frequency $\boldsymbol{\omega}=-q/2m \mathbf{B}$ the time functional is a stationary quantity and GFP restores its validity.

The time derivative of the charge position vector in the rotating frame $\frac{d \mathbf{r} }{d\tau}$ is related to its time derivative in the lab frame $\frac{d \mathbf{r} }{dt}$ by $\frac{d \mathbf{r} }{d\tau} = \frac{d \mathbf{r} }{dt} - \boldsymbol{\omega} \times \mathbf{r}$ \cite{abragam}. By working in the Coulomb gauge, $\mathbf{A}=-\mathbf{r}\times \mathbf{B}/2$, it can be easily shown that $\frac{d \mathbf{r} }{d\tau}=\frac{d \mathbf{r} }{dt}+q\mathbf{A}/m$. The RHS of this equation is nothing but the canonical momentum $\mathbf{P}=m\mathbf{v}+q\mathbf{A}$ divided by the mass of the charge. Let us call the velocities in the rotating frame and the lab frame $\mathbf{u}$ and $\mathbf{v}$ respectively. The time functional in the rotating frame is then  $\mathcal{T}=\int \frac{ds}{u}$. If we parameterize the path in terms of the lab frame time coordinate $t$, the time functional transforms into $\mathcal{T}=\int \frac{v dt}{u}$. Assuming the magnetic field in the $z$ direction and the charge initially having a horizontal velocity, we find that $\mathcal{T}=\int \frac{  \sqrt{\dot x^2 + \dot y^2 } dt  }{\sqrt{(\dot x-\omega y)^2 + (\dot y+\omega x)^2 } }$ where $\omega=|\boldsymbol{\omega}|$. The author has verified numerically that the EL equations are verified for this action and hence $\mathcal{T}$ is a stationary quantity in the rotating frame. 

\section{Discussion} Having demonstrated the validity of the Fermat's principle for quantum Hilbert spaces and the phase space of a system of harmonic oscillators, we need to emphasize that the validity of Eq. \ref{var} is not a trivial consequence of the invariance of $\nu(s)$ along the physical path, but is rather attributed to the underlying dynamics, or more precisely the differential equations governing those dynamics. Not every functional of the form $\int F(s)ds$ is an extremum along the path of constant $F$. On the opposite side, a dynamical system can have an extremum action $\int F(s)ds$ even when $F(s)$ is not a constant of motion as in Eq. \ref{jac}. Moreover, the validity of Eq. \ref{var} in both Hilbert and phase spaces may not be related to the fact that the speed of evolution is directly connected with the energy in both cases since we can easily check that the functional $\int \langle H \rangle ds$ in the first case and $\int H(P_i,Q_i) ds$ in the second case are not stationary along the physical path.  Satisfying  Eq. \ref{var} is not synonymous to the conservation of energy, and we did not have to impose the constraint of constant energy to verify it. Moreover, on the energy manifolds in Hilbert and phase spaces, there is a multitude of trajectories connecting the initial and final states, but not all of them satisfy $\delta  \mathcal{T} =0$.  

While it is true that one can find a variational principle that describes the solution of almost every differential equation, the conceptual significance of the GFP is the link it makes with an established and a well known principle in optics.  The principle advertised in this paper is obviously not as practical as the original Fermat's principle or the Lagrange principle of least action for example since we cannot easily solve Eqs. \ref{fullELPS} and \ref{long} and find the physical path.  However, it illustrates an interesting property of the evolution in Hilbert and phase spaces that we believe is worthy to be highlighted.

 As a leap of faith based on the previous  examples, we claim that GFP will be universally true in any space where the dynamical system evolves with a constant speed. This implies that if the state of the system is projected to a sub-manifold of the full space $ \mathbb{S} $, where a metric is defined and $\nu(s)$ of the projected state is an integral of motion, Fermat's principle will hold in that space as well.  An important exception of this generalization is the motion of a charge in a magnetic field; in this case we found a new frame of reference where GFP is valid in its configuration space.

  The ability to find the manifold or the transformation that renders $\nu(s)$ an integral of motion relies on our ability to find the integrals of motion of the given dynamical system, not a trivial task in many cases. Therefore, the generalized Fermat principle is more likely to be relevant in integrable classical systems than in non-integrable systems.   We recall that the original Fermat's principle is a wave phenomenon. The minimization of the time functional $\mathcal{T}$ occurs along the path which leaves the phase of the light wave stationary with respect to small variations of the path (see \cite{science} and references therein.) The notion of periodic phase variations is found naturally in two types of dynamical systems: completely integrable classical systems and quantum systems. The description of the system in terms of the action-angle variables in the first case and the eigenbasis of the Hamiltonian in the second case offers an intuitive representation of the dynamics as a superposition of {\it waves}. 
  
  On the other hand, even for a completely integrable system, it is not a trivial task to find the metric space where $\nu(s)$ is a constant of motion and it is not even clear whether such a space exists for every integrable system. It is therefore instructive to ask: if we abandon the notion of single trajectories in the classical domain and consider a semiclassical system described by a quasi-probability distribution in the phase space,  can a generalized form of Fermat's principle hold in this representation? It seems plausible to expect interesting phenomena to occur in this case due to the quantum corrections to the classical Liouville's equation or because quantum quasi-probability wavepackets can be thought of as superpositions of many classical trajectories that interfere with each other \cite{martens}. We hope that this discussion will trigger more research efforts into this direction.

 \section{Conclusion}
In conclusion, we have introduced a conjecture that generalizes Fermat's principle in Hilbert and phase spaces of quantum and classical systems respectively and illuminates new aspects of Fermat's surmise of \emph{natural economy}. The generalized Fermat principle  provides  a new geometric variational principle satisfied naturally by the Schr\"odinger and Hamilton's equations of motion in the proper space that has an associated metric distance when the speed of evolution is an integral of motion. This principle may have implications for the protein folding problem, one of the major challenges of biological sciences today \cite{dill}. The mechanism followed by a protein to evolve from its unfolded structure to the native structure that has a minimum  free energy is not fully understood. The puzzle lies in the ultrashort time the protein takes to fold   with respect to the astronomical number of intermediate conformations. Although the folding protein is an open system, we may gain some insight into this problem by searching for the proper space $ \mathbb{S} $ where the generalized Fermat principle is fulfilled.
 
 The author would like to thank Chris Gray for his comments and Boris Fine for several fruitful discussions and useful comments on the manuscript. 

\bibliographystyle{apsrev4-1}
\bibliography{fermat}

\end{document}